\documentclass[a4paper,twocolumn,aps,prl]{revtex4-1}
\usepackage{amsmath}
\usepackage{graphicx}
\usepackage{amssymb}

\newcommand{\ket}[1]{\left|#1\right>}
\newcommand{\bra}[1]{\left<#1\right|}
\newcommand{\abs}[1]{\left|#1\right|}

\newcommand{\avr}[1]{\left<#1\right>}
\newcommand{\bav}[1]{\bigl<#1\bigr>}
\newcommand{\Bav}[1]{\Bigl<#1\Bigr>}

\begin{document}

\title{Emergence of canonical ensembles from pure quantum states}

\author{Jaeyoon Cho}
\author{M. S. Kim}
\affiliation{Institute for Mathematical Sciences, Imperial College London, London SW7 2BW, UK}
\affiliation{QOLS, Blackett Laboratory, Imperial College London, London SW7 2BW, UK}
\affiliation{School of Mathematics and Physics, Queen's University Belfast, University Road, Belfast BT7 1NN, UK}

\date{\today}

\begin{abstract}
We consider a system weakly interacting with a bath as a thermodynamic setting to establish a quantum foundation of statistical physics. It is shown that even if the composite system is initially in an arbitrary nonequilibrium pure quantum state, the unitary dynamics of a generic weak interaction almost always drives the subsystem into the canonical ensemble, in the usual sense of typicality. A crucial step is taken by assuming that the matrix elements of the interaction Hamiltonian have random phases, while their amplitudes are left unrestricted.
\end{abstract}

\maketitle

The concept of typicality has recently gained renewed interest in the context of the quantum foundation of statistical physics~\cite{bocchieri:1959a,*goldstein:2006a,*popescu:2006a,peres:1984a,*tasaki:1998a,*brody:2007a,*reimann:2008a,*linden:2009a,reimann:2007a,*bartsch:2009a}. In particular, the following has been shown~\cite{bocchieri:1959a, goldstein:2006a, popescu:2006a}: Consider a closed quantum system having a very large Hilbert space dimension. One randomly picks a generic pure quantum state among those having a particular energy within an energy shell. It then turns out that the state of a small subsystem with the rest being traced out almost always, i.e., typically, has a canonical distribution. Conceptually, this perspective seems to contrast with the traditional one, in which the canonical ensemble of a subsystem follows when the total system is in the maximum-entropy state, i.e., the microcanonical ensemble~\cite{huang:1987a}. In the new perspective, the entropy of the total system seems to be zero, while its subsystem nevertheless has a large entropy. 

It should be noted here that the above two seemingly different perspectives are equivalent in the following sense: The new perspective implicitly assumes ergodicity as the pure state is taken with respect to the {\em uniform} measure. The initial set of states from which one is taken is thus, due to symmetry, a microcanonical ensemble. As the objective realism is an inherent nature of those statistical descriptions and the microcanonical ensemble is unitarily invariant, whichever basis is chosen, picking one state out of the ensemble, having prior and complete knowledge of the system and hence making the entropy vanish in advance, should not make any difference to the physical consequences. An important point is the fact that when we are given a thermodynamic system, we are already observing only one state among those in the ensemble. Whether we know beforehand which state it was does not matter. What statistical physics asserts is that we indeed need not know such microscopic information because almost all individual states will yield the same statistical quantities as long as their fluctuations vanish in the thermodynamic limit.

It is thus instructive to discriminate between two aspects behind the typicality arguments~\cite{bocchieri:1959a, goldstein:2006a, popescu:2006a}. On one hand, the typicality is generally a property of a large Hilbert space dimension. On the other hand, its actual manifestation (e.g., the resulting subsystem state) depends on the choice of the initial ensemble from which the state is taken. Hence, there remains a fundamental  question yet to be answered: Why do we pick a state with equal priority? Why does the system have no preference to particular states? Recall that the ergodic hypothesis is not a necessary physical consequence that every system should result in, hence, hypothesis. A more relevant question should be when and how the microcanonical ensemble is a plausible representation of a given system, which we shall consider in this paper. Whereas classical concepts such as chaos and mixing support the ergodic hypothesis on a mathematically rigorous footing, there are no stringent counterparts of them in quantum physics, which poses difficulties~\cite{zaslavsky:1981a}. Note, however, that ergodicity and mixing are empirically defined, not derived from first principles, and by nature statistical concepts defined on a measure space. In order to establish a corresponding quantum theory, the question is, as for any statistical argument, how to introduce some statistical nature, i.e., randomness, in a physically plausible manner. In this regard, it is tempting to consider the inherent phase ambiguities in quantum systems as a natural source of randomness. 

In this paper, we consider a closed thermodynamic system, composed of a system S and a bath B, in an arbitrary nonequilibrium pure quantum state with a certain energy, and study how it evolves when a generic weak interaction between S and B, whose meaning is clarified below in terms of phase randomness, is brought in. We show that for almost all interactions system S is driven to the canonical ensemble under some reasonable conditions. To this end, we separate the time-independent Hamiltonian $H$ into the noninteraction part $H_0=H_S+H_B$ with the system (bath) Hamiltonian $H_S$ ($H_B$) and the interaction part $H_1$. We expand the state at time $t$ with the eigenstates $\ket{j}$ of $H_0$ as $\ket{\Psi(t)}=\sum_jc_j(t)\ket{j}$. The transitions between the energy levels are determined by the off-diagonal elements $H_{jk}\equiv\bra{j}H_1\ket{k}$ (we let $H_{jj}=0$). 

Our starting point is to note that, given the inherent complexity and arbitrariness of thermodynamic systems, it is practically reasonable to assume that the phases of $H_{jk}$, which vary sensitively with boundary conditions, positions of interaction, etc., are determined randomly. Let us denote by $\theta_{jk}$ the phase of $H_{jk}$. Our basic assumption is that $\theta_{jk}\in[0,2\pi)$ could be taken independently and randomly with respect to the uniform measure. For later convenience, however, we transform the set of $\theta_{jk}$ into an equivalent set of uniform random variables $\phi_j,\theta_{jk}'\in[0,2\pi)$  for $j,k\neq1$, where $\phi_j=\theta_{j1}$ and $\theta_{jk}'=\theta_{jk}-(\phi_j-\phi_k)$ (let us set $\phi_j=\theta_{jk}'=0$ otherwise). We then take only $\phi_j$ as the random variable while letting $e^{i\theta_{jk}'}$ be absorbed into $H_{jk}$, which is enough to account for the typicality of thermodynamic evolutions. We now have 
\begin{equation}
	H_{jk}\rightarrow e^{i(\phi_j-\phi_k)}H_{jk},\quad\phi_j,\phi_k\in[0,2\pi).
	\label{eq:randomhamil}
\end{equation}
In what follows, we shall denote by $\avr{\cdot}$ the average over the ensemble defined with respect to the random variables $\phi_j$~\cite{:a}. This ensemble reminds us of the random matrix theory~\cite{wigner:1955a,*brody:1981a}. Our assumption is, however, much weaker because except for the phase randomness, no assumptions are made on the statistical distribution of their magnitudes. The formalism below also holds, with slight modification, for real $H_{jk}$ with random signs.

As discussed before, a source of randomness is a necessary element of any statistical argument. In this sense, we simply take the hypothesis of Eq.~\eqref{eq:randomhamil} for granted without a supporting microscopic theory, except for arguing that there is no physical law for an arbitrary system to prefer a particular phase unless specially prepared.  We are just concerned with its consequences when such an ensemble is acceptable as properly describing a given system. Let us take the interaction picture to describe the dynamics in terms of the time evolution operator $U(t,t_{0})=1-i\int_{t_{0}}^tV_I(t')U(t',t_0)dt'$ with $V_I(t)=e^{iH_0t}H_1e^{-iH_0t}$. The coefficients $c_j(t)$ can be written as
\begin{equation}
	c_j(t)=U_{jj}(t,t_{0})c_j(t_{0})+\sum_{k{\not=}j}U_{jk}(t,t_{0})c_k(t_{0})
	\label{eq:evolution}
\end{equation}
with $U_{jk}(t,t_{0})=\bra{j}U(t,t_{0})\ket{k}$. From Eq.~\eqref{eq:randomhamil}, we have
\begin{equation}
	\bav{U_{jk}^{*}(t,t_{0})U_{lm}(t,t_{0})}={\abs{U_{jk}(t,t_{0})}^{2}}\delta_{jl}\delta_{km}
	\label{eq:randomunitary}
\end{equation}
with the Kronecker deltas $\delta_{jk}$ and $\delta_{km}$. Note that the right-hand side is invariant under the phase randomness. Eq.~\eqref{eq:randomunitary} implies that
\begin{equation}
	\avr{c_{j}^*(t)c_{k}(t)}=\bav{\abs{c_{j}(t)}^2}\delta_{jk}
	\label{eq:randomcoeff}
\end{equation}
is satisfied at every $t$. Note that this condition along with
\begin{equation}
	\bav{\abs{c_{j}(t)}^{2}}\simeq\text{constant}
	\label{eq:constantcoeff}
\end{equation}
constitutes the defining character of the ergodic state, or equivalently the microcanonical ensemble. Once condition~\eqref{eq:randomcoeff} is understood, the remaining question is how to {\em derive} condition~\eqref{eq:constantcoeff} from it.

Before proceeding, we clarify three extra (but reasonable) conditions. The first condition is concerning the time scale. Among various time scales characterizing dynamical systems~\cite{casati:1995a}, two time scales are relevant in our context. The first one is the Heisenberg time $\tau_{H}\sim1/\Delta E$, where $\Delta E$ is the characteristic energy level spacing of the system. This time scale is related to the energy-time uncertainty. When $t\ll\tau_{H}$, the system cannot see the discrete nature of its energy spectrum. In this case, if the dynamics is diffusive, it persists as in the classical counterpart system. On the other hand, when $t\gtrsim\tau_{H}$, the system recognizes the discreteness and the dynamics can deviate from the classical counterpart. Dynamical localization~\cite{fishman:1982a}, a dynamical analog of Anderson localization~\cite{anderson:1958a}, is a well-known phenomenon in this regime, demonstrating a stark difference between quantum and classical dynamics. Another relevant time scale is the ergodic time $\tau_{E}$, which estimates the time for the system to wander the entire region of the available phase space. This time scale is largely dependent on the microscopic details, and discussing it quantitatively is out of our scope. Qualitatively, we can argue that the more complex the transition structure is, the faster is the diffusion process, hence the shorter is $\tau_{E}$. Later it becomes evident that one can introduce such a concept as connectivity in the purely classical sense. Here we simply make an inevitable assumption that the transition structure is such that all the energetically allowed levels are indeed connected through finite transitions (in the classical sense) and its time scale meets $\tau_{E}\ll\tau_{H}$. If $\tau_{E}\gtrsim\tau_{H}$, dynamical localization can occur, inhibiting the system from reaching the ergodic state. Given the fact that the total energy increases linearly with the number of particles, while the Hilbert space dimension increases exponentially, for ordinary thermodynamic systems, the energy spectrum is continuous in any physically realistic sense and the Heisenberg time can never be reached. The condition $\tau_{E}\ll\tau_{H}$ is thus automatically met in all such systems as long as the transition structure is as assumed above. At this point, it is worthwhile to compare earlier dynamical considerations in the present context with ours~\cite{peres:1984a,tasaki:1998a,brody:2007a,reimann:2008a, linden:2009a}. Therein, the reason for being able to observe an ensemble behavior was ascribed to the fact that the time average of the relative phase $\avr{e^{-i(E_{j}-E_{k})t}}_{t}$ between two energy eigenstates effectively vanishes in a long time. However, this time averaging effect is relevant after the Heisenberg time. In the thermodynamic systems we consider, this effect cannot be observed.

The second condition is the high dimensionality, which is essential in statistical physics. We quantify this in terms of the effective dimension, or the effective number of involved states, $d_0(t)^{-1}=\sum_j\abs{c_j(t)}^4$, which estimates how many coefficients are nonzero. We assume $d_{0}(t)\gg d_{S}$ is met at every $t$, where $d_{S}$ is the Hilbert space dimension of system S. As our regime is where the total system has an extremely denser energy spectrum than system S, this condition is also automatically met.

The last condition is concerning the interaction between the system and the bath. As the spectrum is a quasicontinuum, we can write the state as $\ket{j}\rightarrow\ket{E_{S},E_{B}}$, where $E_{S}$ and $E_{B}$ denote the energies of system S and bath B, respectively. We assume that $\abs{\bra{E_{S},E_{B}}H_{1}\ket{E_{S}',E_{B}'}}$ is a quasicontinuous function of these energies, or at least of $E_{B}$ and $E_{B}'$ (recall that the bath has a much denser energy spectrum). For example, this is the case in scattering problems, which would describe the dynamics of gases~\cite{bruus:2004a}. In addition, we assume that the interaction is weak. This has a clear analog in statistical physics, wherein the derivation of the canonical ensemble from the microcanonical ensemble relies on the assumption of weak interaction.

We are now in a position to proceed to our main result. From Eq.~\eqref{eq:evolution}, the population in $\ket{{j}}$ can be written as:
\begin{equation}
	p_j(t)=\abs{c_j(t)}^2=p_j^d(t)+[p_j^{f_1}(t)+p_j^{f_{1}*}(t)+p_j^{f_{2}}(t)],
	\label{eq:population}
\end{equation}
where
\begin{align}
	p_j^d(t)=&\abs{U_{jj}(t,t_{0})}^2\abs{c_j(t_{0})}^2+\sum_{k{\not=}j}\abs{U_{jk}(t,t_{0})}^2\abs{c_k(t_{0})}^2,\label{eq:diagonal}\\
	p_j^{f_1}(t)=&U_{jj}(t,t_{0})c_j(t_{0})\sum_{k{\not=}j}U_{jk}^*(t,t_{0})c_k^*(t_{0}),\\
	p_j^{f_2}(t)=&\sum_{k\not=j,l\not=j}U_{jk}(t,t_{0})U_{jl}^*(t,t_{0})c_k(t_{0})c_l^*(t_{0})(1-\delta_{kl}).
\end{align}
Our plan is to calculate the average dynamics, wherein system S asymptotically reaches the canonical ensemble by tracing out over bath B. We then show its fluctuation vanishes, which proves that for almost every {\em individual} dynamics system S reaches the canonical ensemble. 

From Eq.~\eqref{eq:randomunitary} it is easily seen that
\begin{equation}
	\avr{p_{j}(t)}=\avr{p_{j}^{d}(t)}.
	\label{eq:average}
\end{equation}
By defining the transition rate
\begin{equation}
	w_{k\rightarrow j}(t,t_{0})\equiv\frac{d}{dt}\abs{U_{jk}(t,t_{0})}^{2}
\end{equation}
and using $\frac{d}{dt}\abs{U_{jj}(t,t_{0})}^{2}=-\sum_{k\not=j}\frac{d}{dt}\abs{U_{kj}(t,t_{0})}^{2}$, which follows from the unitarity of $U(t,t_{0})$, Eq.~\eqref{eq:average} reads
\begin{equation}
	\begin{split}
	\frac{d}{dt}\avr{p_{j}(t)}=&-\sum_{k\not=j}w_{j\rightarrow k}(t,t_{0})\avr{p_{j}(t_{0})}\\
	&+\sum_{k\not=j}w_{k\rightarrow j}(t,t_{0})\avr{p_{k}(t_{0})}
	\label{eq:transition}
	\end{split}
\end{equation}
As the interaction is weak and the energy spectrum is continuous, the transition rate is given by the Fermi golden rule:
\begin{equation}
	w_{k\rightarrow j}=2\pi\abs{\bra{j}H_{1}\ket{k}}^{2}\delta(E_{j}-E_{k}),
	\label{eq:fermi}
\end{equation}
where it is understood that the density of energy levels $\Gamma(E_{j})$ appears when the summation is taken over the final states around $\ket{j}$, and $E_{j}$ and $E_{k}$ are the corresponding energies~\cite{landau:1981a}. Eq.~\eqref{eq:fermi} can also be calculated up to higher orders using the so-called $T$ matrix, but the result is essentially the same in nature~\cite{bruus:2004a}. 

For a continuous energy spectrum, transition should be treated carefully with respect to a well-defined normalization method~\cite{landau:1981a}. In our case, the Hilbert space can be naturally separated into $d_{S}$ subspaces, $\mathcal{H}=\bigoplus_{\mu=1}^{d_S}\mathcal{H}_\mu$, where $\mathcal{H}_\mu$ is defined by a projector $\ket{\Psi_{\mu}^{S}}\bra{\Psi_{\mu}^{S}}\otimes I^{B}$ with $\ket{\Psi_{\mu}^{S}}$ the $\mu$th eigenstate of $H_{S}$ and $I^{B}$ the identity operator for bath B. We are concerned with the transitions between these subspaces, which represent the transitions in system S of our interest. 

Suppose the total system has energy $E$ within a certain precision, including that due to the energy-time uncertainty. Let us define the population in $\mathcal{H}_{\mu}$:
\begin{equation}
	P_\mu(t)=\sum_{\ket{j}\in\mathcal{H_\mu}}p_j(t).
\end{equation}
We assume $w_{j\rightarrow k}=0$ for $\ket{j},\ket{k}\in\mathcal{H_{\mu}}$, since such transition within a subspace can be included in $H_{B}$. From the definition of $\mathcal{H}_{\mu}$, it can be seen that the density of energy levels of $\mathcal{H}_{\mu}$ is solely determined by that of the bath. Denoting by $E_{\mu}^{S}$ the eigenvalue of $\ket{\Psi_{\mu}^{S}}$, it can be written as $\Gamma_{B}(E-E_{\mu}^{S})$. Note that the transition of Eq.~\eqref{eq:fermi} preserves energy. By suitably summing Eq.~\eqref{eq:transition} and taking into account the energy spectrum being continuous, we can obtain
\begin{equation}
	\begin{split}
	\frac{d}{dt}\avr{P_{\mu}(t)}=&-\sum_{\nu\not=\mu}W_{\mu\rightarrow \nu}\Gamma_{B}(E-E_{\nu}^{S})\avr{P_{\mu}(t_{0})}\\
	&+\sum_{\nu\not=\mu}W_{\nu\rightarrow \mu}\Gamma_{B}(E-E_{\mu}^{S})\avr{P_{\nu}(t_{0})},
	\end{split}
	\label{eq:markov}
\end{equation}
where $W_{\mu\rightarrow\nu}=W_{\nu\rightarrow\mu}$ denotes
\begin{equation}
	W_{\mu\rightarrow\nu}\equiv2\pi\abs{\bra{E_{\nu}^{S},E-E_{\nu}^{S}}H_{1}\ket{E_{\mu}^{S},E-E_{\mu}^{S}}}^{2}
\end{equation}
It is instructive to represent this situation as a transition network of $d_S$ nodes. In particular, as the transition rate is independent of time, the transition structure is visualized as a Markov chain. One can define a population vector $\mathbf{P}(t)$ from $\avr{P_\mu(t)}$ and a transition matrix $\mathbf{T}$ such that
\begin{align}
	T_{\mu\mu}&=1-\sum_{\nu\not=\mu}W_{\mu\rightarrow\nu}\Gamma_B(E-E_\nu^B)\delta t,\\
	T_{\mu\nu}&=W_{\nu\rightarrow\mu}\Gamma_B(E-E_\mu^B)\delta t.
\end{align}
The population is then given by
\begin{equation}
	\mathbf{P}(t_0+n\delta t)=\mathbf{T}^n \mathbf{P}(t_0)
	\label{eq:markovmat}
\end{equation}
for a time interval $\delta t$ chosen to be fairly small, but large enough for the Fermi golden rule to be valid. 

This picture is very useful because the ergodic theory is well established in the Markov chain~\cite{grinstead:1997a}. When every node is reachable from everywhere, the Markov chain is called ergodic or irreducible. In this case, the ergodicity means in the physical context that the time average of the population approaches a steady state in that it is given by $\lim_{n\rightarrow\infty}[\sum_{j=0}^{n}P(j\delta t)]/(n+1)$. An ergodic Markov chain is called regular if some power of the transition matrix has only positive elements. A regular Markov chain has a stronger meaning of ergodicity in that $P(t)$ itself approaches the steady state as $t\rightarrow\infty$. In either case, there is a unique steady state determined by detailed balance. From Eq.~\eqref{eq:markov}, the steady state is found to be
\begin{equation}
	\avr{P_{\mu}(t)}\propto\Gamma_{B}(E-E_{\mu}^{S}).
	\label{eq:steadystate}
\end{equation}
Note that this condition is equivalent to that of the canonical ensemble, where it is understood $\Gamma_B(E)$ increases exponentially with $E$~\cite{huang:1987a}. A similar result can be obtained using the master equation approach~\cite{breuer:2002a,*esposito:2007a}.

Finally we show that the fluctuation of the average dynamics effectively vanishes. Note that $p_{j}^{d}(t)$ in Eq.~\eqref{eq:diagonal} is invariant under the phase randomness for any given state at time $t_{0}$. The off-diagonal contributions $p_{j}^{f_{1}}(t)$ and $p_{j}^{f_{2}}(t)$ in Eq.~\eqref{eq:population} can thus be interpreted as random fluctuations to $p_{j}^{d}(t)$. It is thus enough to show that the variances of $p_{j}^{f_{1}}(t)$ and $p_{j}^{f_{2}}(t)$ vanish in our regime. These are calculated straightforwardly from Eq.~\eqref{eq:randomunitary}. We have
\begin{align}
	\bav{|{p_j^{f_1}(t)}|^2}
	&\!=\!\Bav{\abs{U_{jj}(t,t_{0})}^2\!\abs{c_j(t_{0})}^2\!\sum_{k{\not=}j}\abs{U_{jk}(t,t_{0})}^2\!\abs{c_k(t_{0})}^2} \nonumber \\
	&\!\leq\!\sqrt{\frac{1}{d_0}}\abs{U_{jj}(t,t_{0})}^2\bav{\abs{c_j(t_{0})}^2}.
	\label{eq:off1}
\end{align}
The last line follows from the Cauchy-Schwarz inequality $\bigl(\sum_k\abs{U_{jk}(t,t_{0})}^2\abs{c_k(t_{0})}^2\bigr)^2\leq\bigl(\sum_k\abs{U_{jk}(t,t_{0})}^4\bigr)\bigl(\sum_k\abs{c_k(t_{0})}^4\bigr)$ and the property of unitary operators $\sum_k\abs{U_{jk}(t,t_{0})}^4\leq1$. 
In the same fashion, the variance of $p_j^{f_2}(t)$ is calculated as
\begin{equation}
	\bav{|{p_j^{f_2}(t)}|^2}\leq\sqrt{\frac{1}{d_0}}\sum_{k\not=j}\abs{U_{jk}(t,t_{0})}^2\bav{\abs{c_k(t_{0})}^2}.
	\label{eq:off2}
\end{equation}
By summing Eqs.~\eqref{eq:off1} and \eqref{eq:off2}, we have
\begin{equation}
	\bav{|{\Delta P_\mu(t)}|^2}\leq2\sqrt{\frac{1}{d_0}}\avr{P_\mu(t)}.
	\label{eq:var}
\end{equation}
For any statistical distribution $\{P_{\mu}(t)\}$ of $d_{S}$ variables, this variance is negligible since $d_{0}$ is exponentially larger than $d_{S}$. 

In summary, we have presented a quantum description of the generic thermalization process for a subsystem weakly interacting with a bath. A crucial step was taken by assuming that the matrix elements of the interaction Hamiltonian have random phases as in Eq.~\eqref{eq:randomhamil}, while their amplitudes are left unrestricted. Granting 
three extra conditions stated after Eq.~\eqref{eq:constantcoeff}, we found that the average subsystem dynamics reduces to a Markov chain of Eq.~\eqref{eq:markovmat} whose steady state coincides with a canonical ensemble of Eq.~\eqref{eq:steadystate}. Indeed we found that such an average dynamics is the actual manifestation of almost every individual dynamics as its fluctuation vanishes. This implies that for an arbitrary initial state and a generic unitary dynamics of the whole system as characterized above, the subsystem is almost always led to a canonical ensemble. The irreversibility is not problematic because our time scale is such that the recurrence in quantum dynamics is not observable~\cite{peres:1982a}.

We thank H. Nha and C. Gogolin for helpful discussions. We acknowledge the UK EPSRC for financial support.


\begin{thebibliography}{10}
\bibitem{:a}%
  \BibitemOpen
  \bibinfo {note} {Note that $\phi_j$ take only a small portion of the whole
  randomness and there are many ways of doing that. Even if $\theta_{jk}$ have
  some correlations, one may be able to extract independent random variables
  $\phi_j$ from them. In this sense, our assumption is very general.}%
  \bibAnnoteFile{Stop}{:a}%
\end{thebibliography}
\end{document}